\documentclass[12pt]{iopart}
\usepackage{graphicx}
\usepackage{epstopdf} 
\usepackage{amssymb}
\expandafter\let\csname equation*\endcsname\relax 
\expandafter\let\csname endequation*\endcsname\relax
\usepackage{amsmath}
\usepackage{bm}
\usepackage{multirow} 
\usepackage{multicol}
\usepackage{array}
\usepackage{caption}
\usepackage{tabularx}
\newcolumntype{Y}{>{\centering\arraybackslash}X}

\begin{document} \title{Coherent organization in gene regulation: a study on six networks} 
\author{Ne\c se Aral$^1$ and Alkan Kabak\c c\i o\u glu$^1$} 
\address{$^1$Department of Physics, Ko\c c University, Rumelifeneri Yolu Sar\i yer 34450,
  Istanbul, Turkey} 
\ead{naral@ku.edu.tr}

\begin{abstract}

Structural and dynamical fingerprints of evolutionary optimization in
biological networks are still unclear. We here analyze the dynamics of
genetic regulatory networks responsible for the regulation of
cell-cycle and cell differentiation in three organism or cell types each, and show
that they follow a version of Hebb's rule which we term as
coherence. More precisely, we find that simultaneously expressed genes
with a common target are less likely to conflict at the attractors of
the regulatory dynamics. We then investigate the dependence of
coherence on structural parameters, such as the mean number of inputs
per node and the activatory/repressory interaction ratio, as well
as on dynamically determined quantities, such as the basin size and
the number of expressed genes.

\end{abstract}



\maketitle

\section*{List of abbreviations}
\begin{tabbing}
	GRN \hspace{1cm}\=Gene regulatory network\\
	Th \> T-helper lymphocyte
\end{tabbing}

\section{Introduction}

Gene regulatory networks (GRNs) constitute the backbone of
intracellular functional organization at the molecular scale. These
interaction networks are key to understanding the clockwork operation
of a cell's life cycle~\cite{csikasz2006analysis}, mechanisms of
response to environmental changes~\cite{kashiwagi2006adaptive},
robustness against random fluctuations~\cite{chaves2005robustness,
  garg2009modeling}, the effects of
mutations~\cite{sevim2008chaotic,kaneko2007evolution}, embryonic
development in higher organisms~\cite{zhou2007gene}, etc. Thanks to an
enormous amount of data generated by recent experimental and
computational efforts, we now have access to gene expression profiles
in continuous time and can use them to deduce underlying regulatory
interactions~\cite{kuo2008gene, zhou2003construction}, even speculate
on the evolution of such interactions in historical time
scales~\cite{structure3, brooks2011adaptation}.

Inferring the global GRN of an organism from time-resolved gene
expression data is an ongoing challange~\cite{wang2014review,
  chai2014review}. Therefore, past decade witnessed a growing interest
in identifying key principles that govern the structural organization
of the GRNs~\cite{davidson2010emerging,
  peter2009modularity,thomas2001multistationarity,klemm2005topology}. Viewing
the regulatory network as a collection of functional subunits is a
popular paradigm~\cite{motif1,motif2,motif3,motif4,motif5}, supported
by the observation that certain motifs are frequently encountered in
the regulatory networks of many organisms~\cite{milo2002network}. The
GRN structure is ultimately determined and constrained by the
requirement that the regulatory dynamics delivers a timely production
of necessary proteins. The fact that some of the frequently
encountered motifs promote dynamical stability and robustness to minor
failures is therefore not
surprising~\cite{cinquin2002roles}. Controllability recently emerged
as another defining feature of these complex
systems~\cite{liu2011controllability}, stressing the requirement for a
better understanding of the interplay between the network architecture
and its dynamical behavior. Despite the success of such approaches,
it has been pointed out that there is need to develop new methods
taking different edge signs into
account~\cite{wu2009identification}. Present investigation of coherent
regulation in biological networks is a progress in this direction.

\subsubsection*{Coherent regulation:}

Protein production constitutes about one-half of raw
  material and energy consumption within a growing bacterial cell and
  one-third for a differentiating mammalian
  cell~\cite{li2014quantifying,russell1995energetics,buttgereit1995hierarchy}.
Therefore, it is plausible to ask whether the gene regulation hardware
is wired in a way to achieve the desired functionality with minimal
use of these resources. Considering the structure-dynamics relation
from the perspective of energy efficiency, we propose and provide
evidence that the GRN architecture has been partly shaped to promote
{\it unity of purpose among simultaneously expressed genes sharing a
  common regulatory target}. We refer to such cooperative action of
regulatory genes as ``coherent regulation''~\cite{aral2015coherence}.

The idea that the evolutionary pressure for economy may have shaped
regulatory interactions is not new; for example, it has been exploited
earlier to identify the class of Boolean functions that better model
regulatory dynamics~\cite{ramo2005stability}, or to investigate the
frequency of gene duplication in microbes~\cite{wagner2005energy}. We
claim that, network structures with energy-optimal functionality
should be wired to suppress the expression of ``opposing minority''
regulators. These are transcription factors which, if expressed, would
oppose but not significantly alter the target gene's fate due to
outweighing regulatory pressure favoring the status quo. Networks
where such minority influences are suppressed would display a
disproportionate degree of consensus among simultaneously expressed
regulatory elements acting on a common target, i.e., exhibit coherent
regulation.

Note that, the definition of coherent regulation here is different
from that used in the context of robustness
analysis~\cite{willadsen2008understanding}. Yet another use of similar
terminology appears in the categorization of network
motifs~\cite{alon2006introduction}, where the coherence of a motif is
determined according to the compatibility of alternative directed
paths connecting two nodes. In contrast, the degree of coherence
defined in the present work is not only a function of the interactions
(edges) in the network, but also of the expression states of genes
(nodes).

Investigation of coherence on biological networks requires information
about the regulatory machinery in the cell; in particular the
architecture (say, in the form of a directed graph) and the character
(activation/inhibition) of interactions, as well as a detailed
knowledge of the regulatory dynamics. Dynamical aspects of genetic
regulation have been investigated both on small motifs composed of a
few genes~\cite{alon2007network}, and on larger
networks~\cite{huang2005cell, luscombe2004genomic}. Depending on the
desired resolution, Boolean models~\cite{davidich2008boolean,
  faure2006dynamical, boole1, boole2}, Petri nets~\cite{petri1,
  petri2, petri3}, and differential equation based continuum
models~\cite{dif1, dif2} are the typical approaches employed for this
purpose. A continuum model is indispensable for a high
(time-)resolution study of regulatory dynamics. Simpler Boolean models
have also found a wide area of applicability, mostly in studies where
a coarse characterization of the (quasi-)static stationary states is
acceptable~\cite{li2004yeast,faure2006dynamical}. These approaches
have been successful in modelling the regulation of cell
cycle~\cite{davidich2008boolean,
  chen2004integrative,davidich2008differential}, cell
differentiation~\cite{mendoza1999genetic,albert2003topology,
  gursky2001gap}, circadian clocks~\cite{leloup1998model,
  akman2012digital}, etc. We test our hypothesis on Boolean systems
due to their simplicity and accessibility, although our approach can
be generalized in a straightforward manner to continuum models.

The organization of the paper is as follows: Section~\ref{sec:method}
is a formal introduction to the Boolean network dynamics and coherent
regulation. In Section~\ref{sec:networks}, we introduce six regulatory
networks of different organisms or cell types, for which well-established Boolean
models of regulation were adopted from the literature. Section
~\ref{sec:biological} reports our results which suggest a bias towards
high coherence in these systems, upon comparison with appropriately
constructed random networks.  Section~\ref{sec:results} investigates
structural and dynamical features associated with coherent
regulation. Finally in Section~\ref{sec:conclusion}, we discuss our findings and provide motivation
for further investigation of coherence in complex networks.  Overall,
the current work extends our earlier observation on a single
GRN~\cite{aral2015coherence} to multiple organisms or cell types and suggest that a
bias towards high coherence may be a generic feature of gene
regulation in biological systems.

\section{Computational framework}
\label{sec:method}

\subsection{Time evolution}

Following the standart notation, we describe the expression level of a
gene by a binary variable with values 0 (silent) or 1
(expressed). Therefore, the state of a GRN composed of $n$ nodes at a
given time $t$ is a binary vector ${\boldsymbol \sigma}(t)$ of length
$n$, where $\sigma_i(t)$ is the state of the $i^{th}$ gene ($1\le i \le
n$) at time $t$. Time evolution in a deterministic setting (which
applies to all the models considered here) is described by an
evolution operator
$T$:${\boldsymbol\sigma}(t+1)=T[{\boldsymbol\sigma}(t)]$. Given and
initial condition ${\boldsymbol\sigma}(0)$, the ultimate fate (steady
state) of the system is a cycle of length $q$, where a member state
${\boldsymbol\sigma^*}$ of the cycle satisfies the condition
${\boldsymbol\sigma^*}=T^q[{\boldsymbol \sigma^*}]$. A fixed point is
a trivial cycle with $q=1$.

The rules of dynamics for the biological networks considered in this
study were adopted from the respective
references~\cite{li2004yeast,davidich2008boolean,mendoza1998dynamics,faure2006dynamical,krumsiek2011hierarchical,remy2006logical}. Therefore,
the evolution operator $T$ is different for each network, except the
cell-cycle models of the two variants of yeast for which the same
majority rule was employed. We provide network-specific details of
the GRN structure/dynamics in Section~\ref{sec:networks}.

\subsection{Quantifying coherent regulation}
\label{sec:quant_coh}

\begin{figure}[h] 
  \begin{center} 
    \includegraphics[scale=0.6]{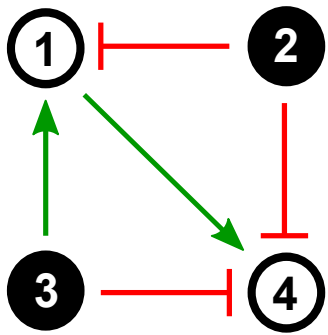} \caption{An example network with
      coherently and incoherently regulated genes. Black and white
      circles indicate active and inactive nodes, respectively. Node 4
      receives coherent input (only repressive) since node 1 is
      inactive. But node 1 itself is incoherently regulated since it
      receives conflicting inputs from nodes
      2\&3.} \label{fig:coherence}
  \end{center} 
\end{figure}

For convenience, let us describe the regulatory influence of a node
$i$ on node $j$ by $c_{ij} \in \{-1,0,1\}$, where $\pm 1$ indicates
positive/negative regulation and $0$ indicates absence of either. In
order to assign a coherence coefficient $\alpha_{c}$ to a GRN, we
first define a single node $i$ of the network to be coherently
regulated in state ${\boldsymbol \sigma}$ if and only if the inputs
from its ``on'' neighbors $j$ ($\sigma_j=1$) are all activatory or all
repressory (see Fig.~\ref{fig:coherence}), i.e., if
\begin{eqnarray} 
  \label{eq:coop_def} 
  \big|\sum_j c_{ij}\sigma_j\big| = \sum_j |c_{ij}|\sigma_j\ .
\end{eqnarray}
The degree of coherence for a state ${\boldsymbol\sigma}$ can then be
measured by the fraction of coherent nodes in that state:
\begin{eqnarray} 
  \label{eq:coh_state}
  \alpha_{c}[{\boldsymbol\sigma}] &=& \frac{1}{n} \sum_i
  \mbox{int}\left(\left|\sum_j c_{ij}\sigma_j\right|\bigg/\sum_j
  \left|c_{ij}\right|\sigma_j\right)\ .
\end{eqnarray} 
A node with no input (both the numerator and the denominator vanish
above) is defined to be perfectly coherent.

We quantify the coherence of the whole network through its steady
states (attractors) corresponding to the fixed points or limit cycles
of the dynamics, in which a cell spends most of its time. Labelling
different steady states with $s$, the global coherence coefficient of
the network is expressed as
\begin{eqnarray} 
  \label{eq:coh_nw}
  \alpha_{c} &=&\sum_s \frac{w_s}{q_s} \sum_{i=1}^{q_s}\alpha_{c}[{\boldsymbol\sigma}^{(s)}_i] 
\end{eqnarray} 
where $q_s$ is the length of the limit cycle $s$ and $w_s$ is a weight
(such as the relative basin size, i.e., the fraction of initial states
that end up in the given attractor) subject to the condition $\sum_s
w_s =1$.

\subsubsection*{Null-model ensembles:} 
Bias towards coherent regulation in a GRN can be assessed by comparing
$\alpha_{c}$ for the given network with the distribution of the same
quantity in a representative ensemble of similar networks. We
construct such a reference ensemble separately for each regulatory
network detailed in the next section.  The ensemble networks were
generated by shuffling the edges of the original GRN sufficiently many
times such that, the resulting network
\begin{itemize}
\item is a connected graph, 
\item strictly conserves the self edges along with the number of
  incoming and outgoing activatory/inhibitory interactons {\it
    separately for each node},
\item statistically has no correlation with the original
  network, except for the local similarities imposed by the two
  constraints above.
\end{itemize}

By construction, each gene in these random ensembles is locally
subject to the same number of repressory and activatory inputs as in
the original network, albeit from possibly different regulatory
partners. Under the same rules of dynamics the fixed points and the
corresponding $\alpha_{c}$ values are generally different for each
random network, since they are determined by the new global
structure. For each ensemble, we generated $10^4$ non-isomorphic
networks and calculated their coherence coefficient both with and
without a basin-size dependent weight ($w_s$) assigned to each
dynamical attractor.

\section{Investigated genetic regulatory networks} 
\label{sec:networks}
\begin{figure}
  \begin{center}
    \includegraphics[width=\textwidth]{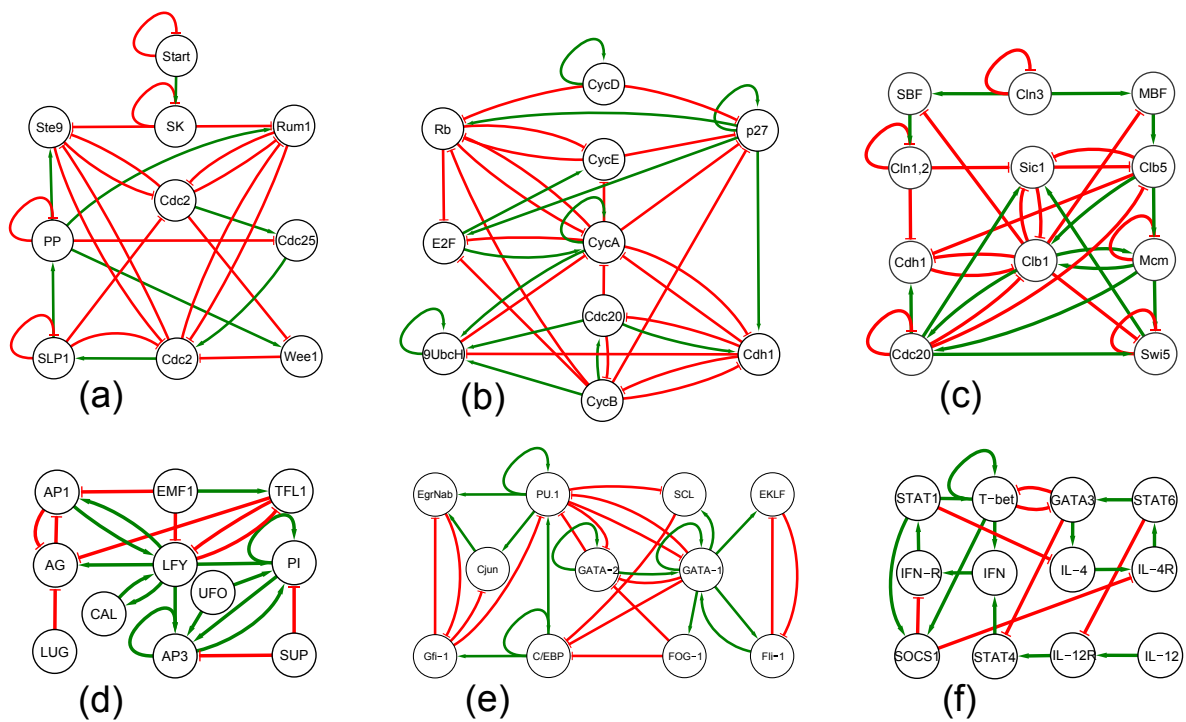}
  \end{center}
  \caption{Gene regulatory networks for (a) Fission yeast cell
    cycle~\cite{davidich2008boolean}; (b) Mammalian cell
    cycle~\cite{faure2006dynamical}; (c) Budding yeast cell
    cycle~\cite{li2004yeast}; (d) \textit{Arabidopsis thaliana} whorl
    differentiation~\cite{mendoza1999genetic}; (e) Myeloid
    differentiation~\cite{krumsiek2011hierarchical}; (f) Th-Lymphocyte
    differentiation~\cite{remy2008minimal}. Here green and red edges indicate activation and inhibition respectively which arise through physical or genetic interactions between the nodes representing genes, proteins, protein complexes or transcription factors.}
  \label{fig:models}
\end{figure}

In this paper, we investigate the degree of coherent regulation in six
GRNs associated with different organisms or cell types: cell-cycle networks in
\textit{Saccharomyces cerevisiae} (budding yeast)~\cite{li2004yeast},
\textit{Scizosaccharomyces pombe} (fission
yeast)~\cite{davidich2008boolean} and
mammals~\cite{faure2006dynamical}, cell-differentiation networks of
\textit{Arabidopsis thaliana} whorls~\cite{mendoza1999genetic}, Th
lymphocyte~\cite{remy2008minimal} and
myeloid progenitors~\cite{krumsiek2011hierarchical}. A graphical description of
each GRN is given in Fig~\ref{fig:models}.  These dynamical models
were chosen from the literature, subject to the criterion that they
reproduce the experimentally observed steady-state expression
profile(s) after truncation to Boolean variables. Below, we give a brief
description of each GRN.\\

{\it Scizosaccharomyces pombe (fission yeast) cell cycle}: The fission
yeast cell-cycle network was modeled by Davidich and
Bornholdt~\cite{davidich2008boolean} as a network with 10 nodes
(Fig.~\ref{fig:models}a). The dynamics is governed by threshold
functions (see Table~\ref{updateyeast}) which yield 12 fixed points
and a fixed cycle. The fixed point with the largest basin matches the
biological $G_1$ phase. \\

{\it Mammalian cell cycle}: Faur\'{e} et al.~\cite{faure2006dynamical}
analysed the regulation dynamics with synchronous, asynchronous, and
mixed updating schemes in this GRN model composed of 10 key regulatory
elements.  Regulation dynamics is given in terms of logical
expressions (see Table~\ref{updatemammalian}) which are determined
according to available experimental evidence for each node. The
resulting dynamical attractors are independent of the updating scheme
and include a fixed point and a limit cycle, in agreement with the
experimental expression data. Note that, the visual depiction of the
GRN given in Ref.\cite{faure2006dynamical} is inconsistent with the
used logic update functions. We here remained faithful to the given
logical expressions, after verifying that they reproduce the reported
steady states. The structure of the GRN consistent with the
interactions in Table~\ref{updatemammalian} is given in
Fig.~\ref{fig:models}b.\\

{\it Saccharomyces cerevisiae (budding yeast) cell cycle}: The model
proposed by Li et al~\cite{li2004yeast} is composed of 11 nodes
(Fig.\ref{fig:models}c). This popular model reproduces the G$_1$
phase as the dominant attractor of a simple Boolean dynamics, as well
as the intermediate phases of cell division. Time evolution is
governed by threshold functions given in Table~\ref{updateyeast},
which yield 7 fixed point attractors. The attractors other than $G_1$
have relatively small basins and, to our knowledge, no clear
biological interpretation.\\

{\it Arabidopsis thaliana whorl differentiation}: Mendoza et
al.~\cite{mendoza1998dynamics} use the network in
Fig.~\ref{fig:models}d in order to model the dynamics of flower
morphogenesis. The interactions in the 11-node model network are again
inferred from experimental data. Different initial conditions evolved
by the proposed rules of dynamics yield 6 point attractors, 4 of which
have a clear biological interpretation.\\

{\it Myeloid differentiation}: A Boolean model was set up by Krumsiek
et al.~\cite{krumsiek2011hierarchical} in order to understand the
mechanisms underlying myeloid differentiation from common myeloid
progenitors to megakaryocytes, erythrocytes, granulocytes and
monocytes. The 11-node model network (Fig.~\ref{fig:models}e) is
composed of relevant transcription factors which evolve (in time)
under separate logical update functions, again inferred from the
available experimental evidence. The dynamics gives rise to 5 point
attractors, where 4 are in agreement with microarray expression
profiles of the mature cell types. It is pointed out that the fifth
attractor cannot be realized during physiological hematopoietic
differentiation.\\

{\it Th-Lymphocyte differentiation}: Remy et
al.~\cite{remy2006logical} proposed this regulatory network model for
the differentiation of T-helper lymphocytes (Th0) cells into Th1 and
Th2 in the vertebrate immune system. Discrete-time evolution of each
node in the model network is governed by node-specific logical rules
given in Table~\ref{updatelymphocyte}. The network structure in
Fig.~\ref{fig:models}f is dominated by activatory interactions.  The
dynamics settles into 3 steady states, in agreeement with the gene
expression levels in the Th0, Th1 and Th2 cells.  \\

\begin{table}
  \begin{center} 
    \begin{tabularx}{0.9\textwidth}{|c*{6}{|Y}|}
      \hline
      \multicolumn{1}{|c|}{{\bf Organism/Cell type}} &\multicolumn{1}{c|}{{\bf Func.}} & \multicolumn{3}{c|}{\bf Network parameters} & \multicolumn{2}{c|}{\bf \# of attr.}\\
      \hline
      & & n & k & p & all & bio.\\ 
      \hline 
      \textit{S. pombe} & c.c. & 10 & 2.3 & 0.35 & 13 & 1 \\ 
      \hline 
      Mammalian cell & c.c.& 10 & 3.1 & 0.32 & 2 & 2 \\ 
      \hline 
      \textit{S. cerevisiae} & c.c. & 11 & 2.64 & 0.51 & 6 & 1  \\ 
      \hline 
      \textit{A. thaliana} whorls & diff. & 11 & 2 & 0.55 & 6 & 4 \\ 
      \hline
      Myeloid progenitor& diff. & 11 & 2.36 & 0.42 & 5 & 4 \\ 
      \hline 
      Th-lymphocyte & diff. & 12 & 1.67 & 0.65 & 6 & 2 \\ 
      \hline 
    \end{tabularx} 
    \caption{Structural/dynamical features associated with the six
      regulatory networks adopted from the literature. ``c.c'' and
      ``diff.'' refer to regulation of cell cycle and differentiation
      respectively. Last two columns exclude the trivial point
      attractor ${\boldsymbol\sigma}=0$.}
      \label{table:grn_param} 
  \end{center} 
\end{table}

It was shown earlier that structural parameters such as the mean
number of incoming edges per node ($k$) and the fraction of
up-regulating interactions ($p$) in the network are important
determinants of global coherence, while the size of the network for
fixed $(k,p)$ only weakly influences
$\alpha_c$~\cite{aral2015coherence}. Observed values of $(n,k,p)$ for
each network are listed in Table~\ref{table:grn_param}. For the GRNs
we consider here, $k$ varies within $[1.67,3.1]$, while $p$ changes in
the interval $[0.32,0.65]$ (self-edges are excluded).  All models have
roughly the same network size ($\sim 11$ nodes).

\section{Coherence in biological networks} 
\label{sec:biological}

\begin{figure}
  \begin{center}
    \includegraphics[width=\textwidth]{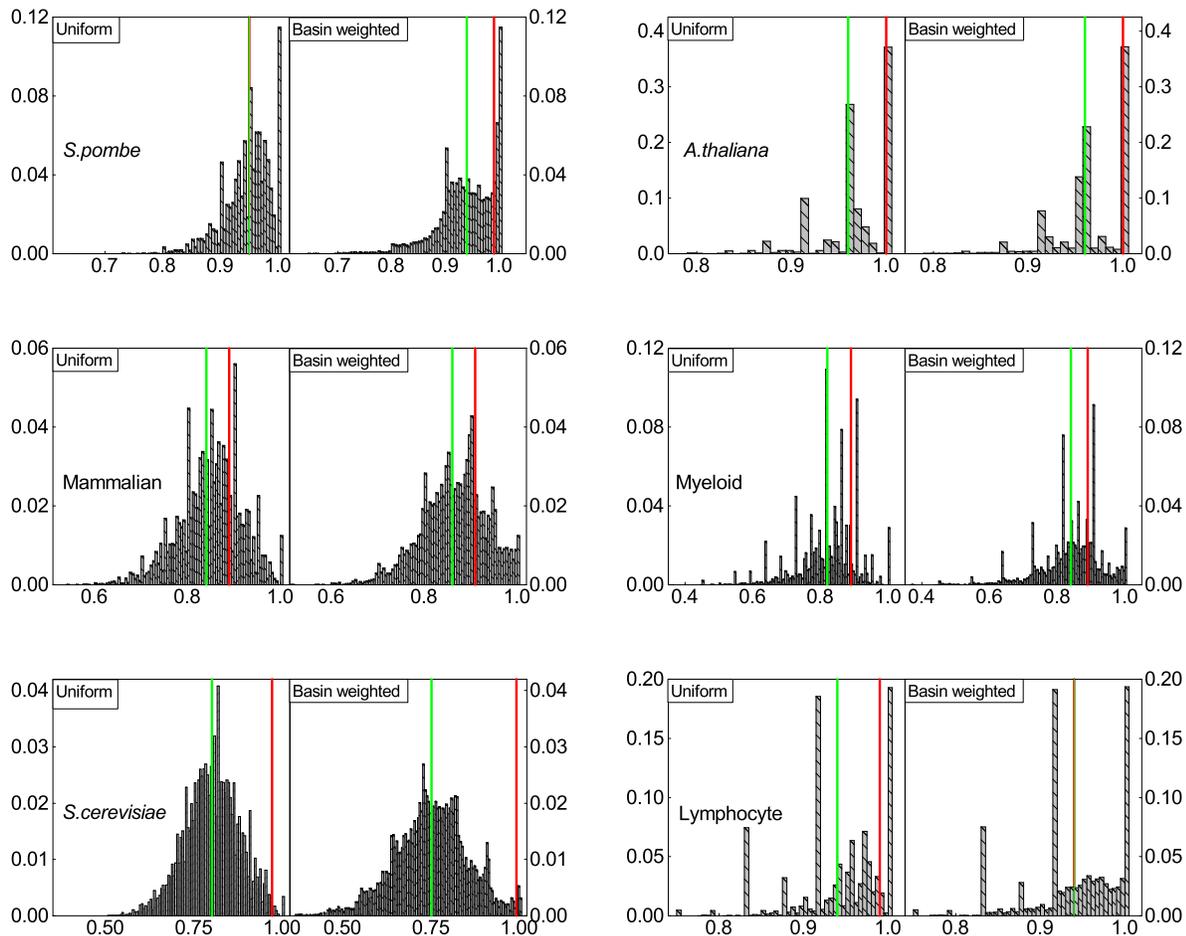}
    \caption{Degree of coherent regulation for biological {\it vs}
      random networks. The histograms of $\alpha_c$ for random ensembles of $10^4$ networks having similar structure is shown with their mean values (green line). The red line indicates the $\alpha_c$ of the biological networks, whose exact values can be seen on Table \ref{table:coherence}.  }
    \label{fig:barchart}
  \end{center}
\end{figure}

Table~\ref{table:coherence} and Fig.~\ref{fig:barchart} summarize the
outcome of our coherence analysis on the GRNs depicted in
Fig.~\ref{fig:models} and listed in section~\ref{sec:networks}. We
separately calculated the degree of coherence of each GRN over the
full set of fixed points/cycles ($\alpha_c$) and over the biologically
relevant subset ($\alpha^{bio}_c$), both adopted from respective
references. We compared them with the mean ($\alpha^{rand}_c$)
obtained from the associated ensemble of random networks described in
Section~\ref{sec:quant_coh}.

\begin{table}[h]
  \begin{center} \small 
    \begin{tabular}{|c||c|c|c||c|c|c|}
      \hline
	  {\bf Organism/Cell type}&\multicolumn{3}{c||}{\bf $\alpha_c$ (uniformly weighted)}&\multicolumn{3}{c|}{\bf $\alpha_c$
	    (basin-size weighted)} \\
	  \hline 
	  & $\alpha_c$ & $\alpha_c^{bio}$ & $\alpha_c^{rand}$ & $\alpha_c$ & $\alpha_c^{bio}$ &$\alpha_c^{rand}$ \\ 
	  \hline
	  \textit{S. pombe}& 0.95  & 1& 0.94 $\pm$ 0.04  & 0.99 & 1 & 0.93$\pm$ 0.05\\ 
	  \hline 
	  Mammalian cell & 0.88  & 0.88& 0.85 $\pm$ 0.07 & 0.88 &0.88 & 0.86 $\pm$ 0.07 \\ 
	  \hline 
	  \textit{S. cerevisiae} & 0.97  & 1 & 0.80 $\pm$ 0.08 & 0.99 & 1 & 0.75$\pm$ 0.10\\ 
	  \hline 
	  \textit{A. thaliana} whorls & 1  & 1& 0.97 $\pm$ 0.04 & 1 & 1 & 0.96 $\pm$ 0.04 \\ 
	  \hline
	  Myeloid progenitor& 0.89  & 0.89& 0.82 $\pm$ 0.09 & 0.90 & 0.88 & 0.84 $\pm$ 0.09 \\ 
	  \hline 
	  Th-lymphocyte & 0.99  & 0.96& 0.94 $\pm$ 0.05 & 0.94 & 0.92 & 0.94 $\pm$ 0.05\\ \hline 
    \end{tabular} 
    \caption{Degrees of coherence associated with the regulatory
      networks shown in Fig.~\ref{fig:models}. $\alpha_c^{all}$,
      $\alpha_c^{bio}$, $\alpha_c^{rand}$, denote the coherence
      value calculated by using Eq.(\ref{eq:coh_nw}) over all attractors of
      the model dynamics, only the attractors observed in the wild
      type, and all networks in the corresponding randomized ensemble,
      respectively.}
    \label{table:coherence}
  \end{center} 
\end{table}

Our central observation is that, despite the variability in $k$ and
$p$, all biological networks are more coherent than random networks
constructed with similar structural parameters.
Fig.~\ref{fig:barchart} also shows the distribution of $\alpha_c$ over
the random ensembles for a better judgment of the bias towards
coherence. The difference in coherence between the biological network
and the random ensemble is visible but within acceptable bounds for
most isolated examples. However, the fact that all display a positive
deviation from the respective ensemble means suggests an overall
preference towards coherence. We quantified the statistically
significance of this bias by using Fisher's
method~\cite{fisher1925statistical} on the present data, which yields
a likelihood of $0.044$ and $0.075$ for a chance encounter of a this
much or larger deviation in uniformly and basin-weighted cases,
respectively.  The degree of selection is appreciable, considering
that the compared ensemble networks have not only the same number of
nodes, the same number of interactions, and the same proportion of
repressory/activatory regulation globally, but also the same local
pattern of incoming and outgoing edges for each node.

\section{Coherence {\em vs} structural/dynamical network properties }
\label{sec:results}

The bias observed in biological regulatory networks above provides
motivation to investigate the relationship between coherence and other
architectural or dynamical determinants of a network. Identifying such
connections could help one recognize coherent systems from certain
telltale patterns instead of requiring detailed information about
their function, and/or design them by means of simple guiding
principles.

\subsection*{Edge number $\&$ type:}
\begin{figure} 
  \begin{center}
    \includegraphics[scale=1]{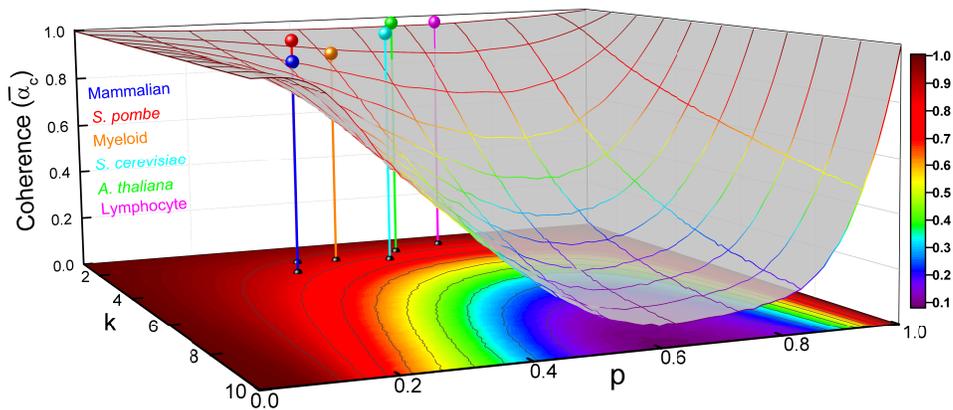}
    \caption{Dependence of coherence value ($\overline{\alpha}_{c}$) on the network
      parameters ($k,p$) with $n=11$. As $k$ increases from 1 to 10, the $p$ value where $\overline{\alpha}_{c}$ is minumum for the fixed $k$ shifts from 0.72 to 0.6. Coherence values corresponding to the
      biological networks are shown by spheres.}
    \label{fig:Nkp_surface_origin}
  \end{center}
\end{figure}

We first investigated the dependence of average coherence
$\overline{\alpha}_{c}$ on the number of incoming/outgoing edges per
node, $k$, and the fraction of activatory interactions, $p$. To this
end, we simulated the majority-rule dynamics in
Ref.~\cite{li2004yeast} on different $(k,p)$ pairs equally spaced in
the rectangle $0.909\le k\le 10\ \otimes\ 0\le p \le 1$, with
$n=11$. The result is shown in Fig.~\ref{fig:Nkp_surface_origin}. Each
data point in Fig.~\ref{fig:Nkp_surface_origin} is an average over
$10^3$ random networks which are generated by random shuffling of
edges. No constraint was imposed on the shuffling process, apart from
the requirement of connectedness (every node is accessible from every
other node in the undirected network).

One observes that, independent of the average connectivity $k$, random
networks display minimum coherence when approximately two-thirds
($60-72\%$) of the interactions are activatory. An intuitive
understanding of this behavior was proposed in
Ref.~\cite{aral2015coherence} where a subset of the results in
Fig.~\ref{fig:Nkp_surface_origin} (for a single $k$ value) was
reported. 

Coordinates corresponding to the studied biological networks are shown
by colored spheres on the same figure. It is interesting that all of
them are situated on the low-$p$ slope of the minimum coherence
valley.

\subsection*{Number of active genes:}
\begin{figure}[h]
  \begin{center} 
    \includegraphics[width=0.8\textwidth]{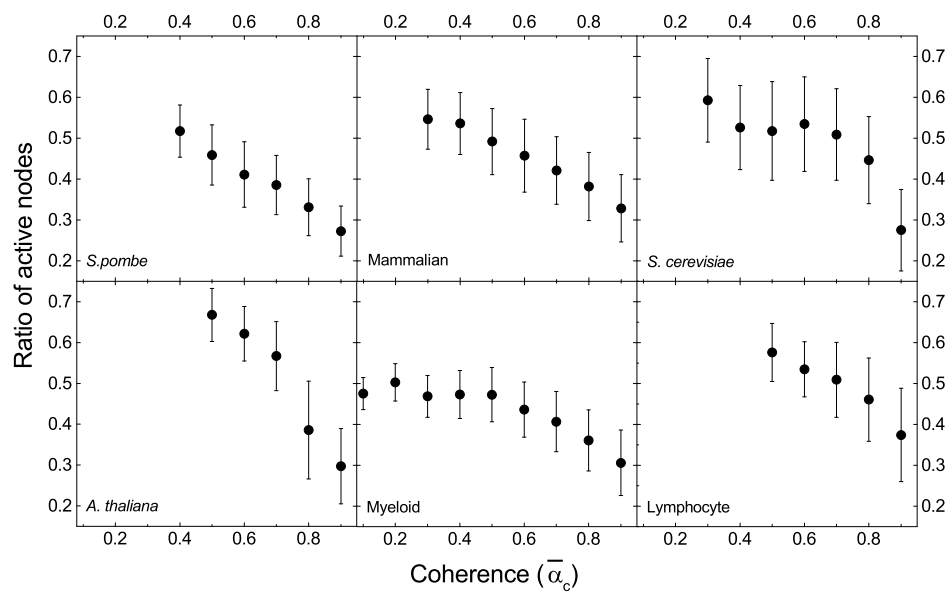}
    \caption{Average ratio of active nodes in the point-attractors vs.
      average coherence in ensembles of model networks. The average is calculated over $10^4$ networks chosen randomly from the null-model ensemble described in section \ref{sec:quant_coh}}
    \label{fig:noofactive}
  \end{center}
\end{figure}

We next asked if the number of active genes $(n_a)$ at a
fixed point/cycle is a determining factor for coherence. After all,
coherence reflects the harmony between the regulatory messages
originating from these genes. To this end, we considered the
attractors found within the ensembles generated by shuffling the edges
of each GRN in Fig.~\ref{fig:models}. We grouped them according to
$n_a$ and calculated the mean coherence within each group. The results
in Fig.~\ref{fig:noofactive} show that attractors with higher number
of active nodes are generally less coherent. This behavior may be
understood by the intuitive fact that it is more difficult to reach
consensus in a large group than in a small one. We remind that, this
relationship is valid only within the restricted ensembles generated
by the shuffling process described in Section~\ref{sec:method}. One
can not speak of a monotonic dependence of coherence on $n_a$ if, for
example, the difference in $n_a$ between two networks is due to
different $p$ values. This is evident from the fact that $\bar{n}_a
(p)$ calculated over the network ensembles used in
Fig.~\ref{fig:Nkp_surface_origin} trivially increases with $p$ while
$\overline\alpha_c (p)$ is nonmonotonic for any fixed $k$.

\subsection*{Basin size:}
\begin{figure} 
  \begin{center} \includegraphics[width=0.8\textwidth]{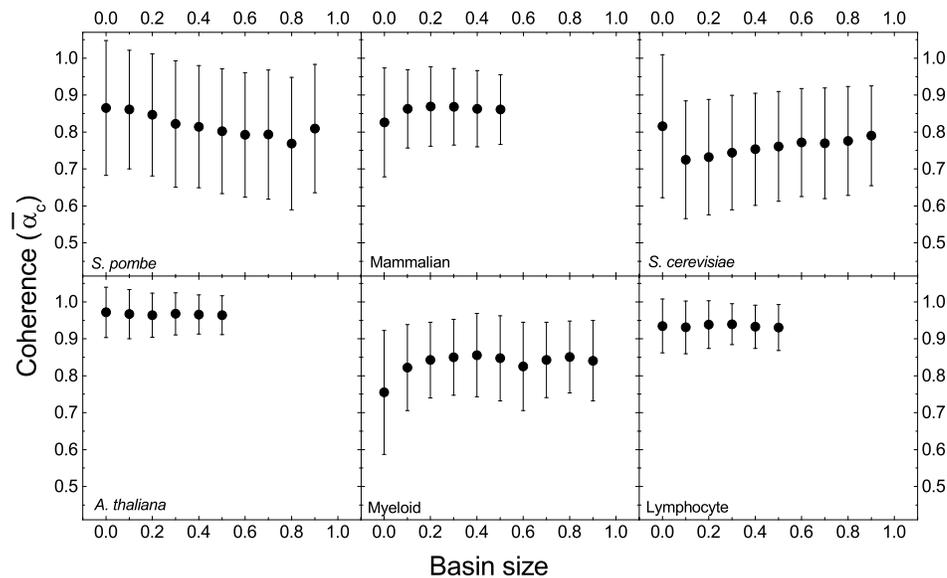}
    \caption{Average coherence vs. average basin size for the
      attractors (point or cycles) of $10^4$ networks with similar
      structure. Note that, proposed rules of dynamics in some models
      appear to prohibit appearance of excessively dominant
      attractors.}
    \label{fig:coherence_vs_basin} 
  \end{center}
\end{figure}

It is implicit from Table~\ref{table:coherence} and
Fig.~\ref{fig:barchart} that, coherent attractors are not always those
with larger basins. We explicitly examined the variation in coherence
as a function of the relative basin size of the attractors.  The
results (Fig.~\ref{fig:coherence_vs_basin}), obtained over the
networks in the randomized ensembles of each GRN separately, suggest
no consistent relation between the basin size and coherence. Likewise,
biologically relevant attractors~\cite{davidich2008boolean,
  faure2006dynamical,li2004yeast,mendoza1998dynamics,krumsiek2011hierarchical,remy2006logical}
of GRNs in Fig.~\ref{fig:models} are not always those with the largest
basins. This is hardly surprising, since the initial state prior to
differentiation or cell division is never determined randomly.

\section{Conclusion}
\label{sec:conclusion}
We investigated the degree of coherent regulation in several
regulatory networks responsible for cell cycle and differentiation in
various organisms or cell types, by means of a recently proposed measure of
coherence. We found that, even though most networks are moderately
more coherent than expected (in reference to architecturally similar
network ensembles), cumulatively there is a statistically meaningful
bias towards coherence.  Our findings lend support to the thesis that
pressure for coherent regulation is one of the factors driving the
evolution of GRNs. It is not difficult to imagine a Hebbian-like
mechanism~\cite{morris1999hebb} in the given context, where regulatory
interactions incompatible with the required expression profile get
eliminated over time. The fact that transcriptional regulatory
networks have been observed to have the fastest mutation rate among
various biological networks~\cite{shou2011measuring} suggests that
their evolutionary dynamics may be sufficiently fluid for effective
Hebbian selection.

We also showed, on randomly generated model GRNs, that coherence is
harder to achieve with increasing number of interactions per node, and
with an inhibitory interaction ratio around $1/3$.  Furthermore,
coherent networks typically involve a smaller number of active genes
at their steady states, compared to arbitrary networks with similar
edge composition and local connectivity. It could be interesting to
focus on networks following the identical expression pattern in time
as, say, the yeast cell-cycle (see, e.g., Ref.~\cite{lau2007function})
and check if those with higher coherence are more similar to their
biological counterpart or encompass other desirable properties such as
higher robustness or better controllability. Finally, from an
architectural perspective, coherence can be viewed as a design choice
for any natural or artificial network where inhibitory and activatory
interactions coexist. One might then ask how to build a coherent
system from scratch, or how to enhance coherence in an existing system
with minimal intervention. We hope our findings to trigger further
theoretical investigations in these directions.

\clearpage	
\section{Appendix}
\label{sec:appendix}

Below, we list the rules of regulatory dynamics for each GRN model
adopted from the literature.

\begin{table}[h]
  \begin{tabular}{|l|l|l|}
    \hline 
    Network & Gene & Boolean Update Function\\
    \hline
    \multirow{11}{*}{\parbox[c]{2cm}{Myeloid differentiation}}& GATA-2&\small{$ \textnormal{GATA-2} \wedge
      \overline{(\textnormal{GATA-1} \wedge \textnormal{FOG-1})} \wedge \overline{\textnormal{PU.1}}$}\\ 
    \cline{2-3} 
    & GATA-1& \small{$(\textnormal{GATA-1} \vee \textnormal{GATA-2} \vee \textnormal{Fli-1}) \wedge \overline{\textnormal{PU.1}}$}\\
    \cline{2-3} 
    & FOG-1 & \small{GATA-1}\\ \cline{2-3} & EKLF & \small{$\textnormal{GATA-1} \wedge
      \overline{\textnormal{Fli-1}}$}\\ \cline{2-3} & Fli-1 & \small{$\textnormal{GATA-1} \wedge
      \overline{\textnormal{EKLF}}$}\\ \cline{2-3} & SCL & \small{$\textnormal{GATA-1} \wedge \overline{\textnormal{PU.1}}$}\\
    \cline{2-3} 
    & C/EBP$_{\alpha}$ & \small{$\textnormal{C/EBP}_{\alpha} \wedge \overline{(\textnormal{GATA-1}\wedge
	\textnormal{FOG-1} \wedge \textnormal{SCL})}$}\\ 
    \cline{2-3} 
    & PU.1 & \small{$(\textnormal{C/EBP}_{\alpha} \vee \textnormal{PU.1}) \wedge \overline{(\textnormal{GATA-1} \vee \textnormal{GATA-2})}$}\\ 
    \cline{2-3} 
    & cJun &\small{$\textnormal{PU.1} \wedge \overline{\textnormal{Gfi-1}}$}\\ 
    \cline{2-3} 
    & EgrNab & \small{$(\textnormal{PU.1} \wedge \textnormal{cJun}) \wedge \overline{\textnormal{Gfi-1}}$}\\ 
    \cline{2-3} 
    & Gfi-1 & \small{$\textnormal{C/EBP}_{\alpha} \wedge \overline{\textnormal{EgrNab}}$}\\ 
    \hline 
  \end{tabular} 
  \caption{Update functions for myeloid differentiation.} 
  \label{updatemyeloid} 
\end{table}

\begin{table}[h] 
  \begin{tabular}{|l|l|l|} 
    \hline Network & Gene & Boolean Update Functiont\\ 
    \hline
    \multirow{10}{*}{\parbox[c]{2cm}{Mammalian cell cycle}}& CycD&$ CycD$\\
    \cline{2-3} 
    & Rb &\parbox{10cm}{$(\overline{CycD} \wedge \overline {CycE} \wedge \overline{CycA} \wedge \overline{CycB}) \vee (p27 \wedge
      \overline{CycD} \wedge \overline{CycB})$} \\ 
    \cline{2-3} 
    & E2F &\parbox{10cm}{$(\overline{Rb} \wedge \overline{CycA}
      \wedge \overline{CycB}) \vee (p27 \wedge \overline{Rb} \wedge \overline{CycB})$} \\ \cline{2-3} & CycE & $E2F  \wedge
    \overline{{Rb}}$\\ 
    \cline{2-3} 
    & CycA & \parbox{10cm}{$ (E2F \wedge \overline{Rb} \wedge \overline{Cdc20} \wedge
      \overline{(Cdh1 \wedge Ubc)})\\
      \vee (CycA \wedge \overline{Rb} \wedge \overline{Cdc20} \wedge \overline{(Cdh1 \wedge
	Ubc)})$}\\
    \cline{2-3} 
    & p27 &\parbox{10cm}{\scriptsize{$(\overline{CycD} \wedge \overline{CycE} \wedge \overline{CycA}
	\wedge \overline{CycB}) \vee (p27 \wedge \overline{(CycE \wedge CycA)} \wedge \overline{CycB} \wedge \overline{CycD})$}}
    \\ 
    \cline{2-3} 
    & Cdc20$_{\alpha}$ & $CycB$  \\
    \cline{2-3} 
    & Cdh1 & \parbox{10cm}{$(\overline{CycA} \wedge
      \overline{CycB}) \vee Cdc20 \vee (p27 \wedge \overline{CycB})$}\\ 
    \cline{2-3} 
    & UbcH10 &\parbox{10cm}{$\overline{Cdh1}
      \vee (Cdh1 \wedge Ubc \wedge (Cdc20 \vee CycA \vee CycB))$}  \\ 
    \cline{2-3} 
    & CycB &$\overline{Cdc20} \wedge \overline{Cdh1}$ \\ 
    \cline{2-3}
    \hline 
  \end{tabular}
  \caption{Update functions for mammalian cell cycle.}
  \label{updatemammalian} 
\end{table}

\begin{center}
  \begin{table}[h!] 
    \begin{tabular}{|p{2.5cm}|c|l|l|} 
      \hline Network & Node $i$ & Gene & $\text{Update
        Function}^{\dagger}$\\ \hline
      \multirow{11}{*}{\parbox[c]{2.5cm}{\textit{S. cerevisiae}\\ cell
          cycle}}& 1 & Cln3 & \multirow{5}{*}{ $S_i(t+1)=
        \left\{\begin{array}{ll} 1, & \mbox{$I>0$}\\ 0, & \mbox{$I
          \leq0$}\\ \end{array}\right.$} \\ & 2 & Cln1,2 &\\ & 3 &
      Cdc20 &\\ & 4 & Mcm &\\ & 5 & Swi5 &\\ \cline{2-4} & 6 & SBF &
      \multirow{6}{*}{ $S_i(t+1)= \left\{\begin{array}{ll} 1, &
        \mbox{$I>0$}\\ 0, & \mbox{$I<0$}\\ S_i(t), &
        \mbox{$I=0$}\\ \end{array} \right.$}\\ & 7 & MBF &\\ & 8 &
      Sic1 &\\ & 9 & Clb5 &\\ & 10 & Cdh1 &\\ & 11 & Clb1
      &\\ \cline{1-4}
      \multirow{9}{*}{\parbox[r]{2.5cm}{\textit{S. pombe}\\ cell
          cycle}} & 1 & SK & \multirow{3}{*}{ $S_i(t+1)=
        \left\{\begin{array}{ll} 1, & \mbox{$I>0$}\\ 0, &
        \mbox{$I\leq0$}\\ \end{array} \right.$} \\ & 2 & SLP &\\ & 3 &
      PP &\\ \cline{2-4} & 4 & Ste9 & \multirow{6}{*}{ $S_i(t+1)=
        \left\{\begin{array}{ll} 1, & \mbox{$I>\theta_i$}\\ 0, &
        \mbox{$I<\theta_i$}\\ S_i(t), &
        \mbox{$I=\theta_i,$}\\ \end{array} \right.$}\\ & 5 & Rum1
      &\\ & 6 & Cdc2 &\\ & 7 & Cdc2* &\\ & 8 & Wee1 &\\ & 9 & Cdc25
      &\\ \cline{1-4}
      \multirow{12}{*}{\parbox[c]{2.5cm}{\textit{A. thaliana}\\ whorl\\differentiation}}
      & 1 & EMF1 & \multirow{12}{*}{ $S_i(t+1)=
        \left\{\begin{array}{ll} 1, & \mbox{$I>0$}\\ 0, &
        \mbox{$I\leq0$}\\ \end{array} \right.$} \\ & 2 & TFL1&\\ & 3 &
      LFY&\\ & 4& AP1&\\ & 5 & CAL&\\ & 6& LUG&\\ & 7& UFO&\\ & 8&
      BFU&\\ & 9& AG&\\ & 10 & AP3&\\ & 11 & PI &\\ & 12 &
      SUP&\\ \hline \multicolumn{4}{|l|}{\parbox[c]{\textwidth}{In the
          update functions $I=\sum_{j=1}^N A_{ij} S_i (t)$ where $A$
          is the weighted adjacency matrix and $\theta$'s indicate the
          thereshold values for the nodes. For detailed explanations
          for these values one can refer to
          \cite{li2004yeast,davidich2008boolean,mendoza1998dynamics}.}}
      \\ \hline
    \end{tabular} 
    \caption{Update rules of the regulatory dynamics for the budding
      yeast ({\em Saccharomyces cerevisiae}), the fission yeast ({\it
        Saccharomyces pombe}), and the flower {\em Arabidopsis thaliana}.}
    \label{updateyeast}
  \end{table}
\end{center}

\begin{table}[h!]
  \begin{tabular}{|l|l|l|} 
    \hline Network & Gene & Boolean Update Function$^\dagger$\\
    \hline
    \multirow{12}{*}{\parbox[c]{3cm}{Lymphocyte Differentiation}}& IFN-$\gamma$& $K_1(9), K_1(11), K_1(9,11)$ \\ 
    \cline{2-3}
    & IL-4& \parbox{6cm}{$K_2(12)$}  \\ 
    \cline{2-3} 
    & IL-12&\parbox{6cm}{} \\ 
    \cline{2-3} 
    & IFN-$\gamma$R&$K_4(1)$ \\
    \cline{2-3} 
    & IL-4R& \parbox{6cm}{$K_5(2)$}\\ 
    \cline{2-3} 
    & IL-12R&\parbox{6cm}{$K_6(3)$} \\ 
    \cline{2-3} 
    & STAT1&$K_7(4)$\\ 
    \cline{2-3} 
    & STAT6 & \parbox{6cm}{$K_8(5)$}\\
    \cline{2-3}
    & STAT4&\parbox{6cm}{$K_9(6)$}  \\ 
    \cline{2-3} 
    & SOCS1&$K_{10}(7),K_{10}(11),K_{10}(7,11)$ \\ 
    \cline{2-3} & T-bet&$K_{11}(7),K_{11}(11),K_{11}(7,11)$ \\ 
    \cline{2-3} 
    & GATA-3&$K_{12}(8)$ \\
    \hline 
    \multicolumn{3}{|l|}{\parbox{\textwidth}{$K_i(\{j\})$ is the set of nodes $j$ that when
	simultaneously active turn the node $i$ on. For details one can refer to \cite{remy2006logical}.}}\\ 
    \hline
  \end{tabular} 
  \caption{Update functions for lymphocyte differentiation.} 
  \label{updatelymphocyte}
\end{table}
\clearpage	
\section*{References} 
\bibliographystyle{iopart-num} 
\bibliography{Coherent_organization_bib}

\providecommand{\newblock}{}
\begin{thebibliography}{10}
\expandafter\ifx\csname url\endcsname\relax
  \def\url#1{{\tt #1}}\fi
\expandafter\ifx\csname urlprefix\endcsname\relax\def\urlprefix{URL }\fi
\providecommand{\eprint}[2][]{\url{#2}}

\bibitem{csikasz2006analysis}
Csik{\'a}sz-Nagy A, Battogtokh D, Chen K~C, Nov{\'a}k B and Tyson J~J 2006 {\em
  Biophysical journal\/} {\bf 90} 4361--4379

\bibitem{kashiwagi2006adaptive}
Kashiwagi A, Urabe I, Kaneko K and Yomo T 2006 {\em PloS one\/} {\bf 1} e49

\bibitem{chaves2005robustness}
Chaves M, Albert R and Sontag E~D 2005 {\em Journal of theoretical biology\/}
  {\bf 235} 431--449

\bibitem{garg2009modeling}
Garg A, Mohanram K, Di~Cara A, De~Micheli G and Xenarios I 2009 {\em
  Bioinformatics\/} {\bf 25} i101--i109

\bibitem{sevim2008chaotic}
Sevim V and Rikvold P~A 2008 {\em Journal of theoretical biology\/} {\bf 253}
  323--332

\bibitem{kaneko2007evolution}
Kaneko K 2007 {\em PLoS One\/} {\bf 2} e434

\bibitem{zhou2007gene}
Zhou Q, Chipperfield H, Melton D~A and Wong W~H 2007 {\em Proceedings of the
  National Academy of Sciences\/} {\bf 104} 16438--16443

\bibitem{kuo2008gene}
Kuo-Ching L and Xiaodong W 2008 {\em EURASIP Journal on Bioinformatics and
  Systems Biology\/} {\bf 2008}

\bibitem{zhou2003construction}
Zhou X, Wang X and Dougherty E~R 2003 {\em Signal Processing\/} {\bf 83}
  745--761

\bibitem{structure3}
Kaneko K 2007 {\em PLoS One\/} {\bf 2} e434

\bibitem{brooks2011adaptation}
Brooks A~N, Turkarslan S, Beer K~D, Yin~Lo F and Baliga N~S 2011 {\em Wiley
  Interdisciplinary Reviews: Systems Biology and Medicine\/} {\bf 3} 544--561

\bibitem{wang2014review}
Wang Y and Huang H 2014 {\em Journal of theoretical biology\/}

\bibitem{chai2014review}
Chai L~E, Loh S~K, Low S~T, Mohamad M~S, Deris S and Zakaria Z 2014 {\em
  Computers in biology and medicine\/} {\bf 48} 55--65

\bibitem{davidson2010emerging}
Davidson E~H 2010 {\em Nature\/} {\bf 468} 911--920

\bibitem{peter2009modularity}
Peter I~S and Davidson E~H 2009 {\em FEBS letters\/} {\bf 583} 3948--3958

\bibitem{thomas2001multistationarity}
Thomas R and Kaufman M 2001 {\em Chaos: An Interdisciplinary Journal of
  Nonlinear Science\/} {\bf 11} 170--179

\bibitem{klemm2005topology}
Klemm K and Bornholdt S 2005 {\em Proceedings of the National Academy of
  Sciences of the United States of America\/} {\bf 102} 18414--18419

\bibitem{motif1}
Thieffry D 2007 {\em Briefings in bioinformatics\/} {\bf 8} 220--225

\bibitem{motif2}
Thomas R, Thieffry D and Kaufman M 1995 {\em Bulletin of mathematical
  biology\/} {\bf 57} 247--276

\bibitem{motif3}
Kaufman M and Thomas R 2003 {\em Comptes rendus biologies\/} {\bf 326} 205--214

\bibitem{motif4}
Burda Z, Krzywicki A, Martin O and Zagorski M 2011 {\em Proceedings of the
  National Academy of Sciences\/} {\bf 108} 17263--17268

\bibitem{motif5}
Milo R, Shen-Orr S, Itzkovitz S, Kashtan N, Chklovskii D and Alon U 2002 {\em
  Science Signalling\/} {\bf 298} 824

\bibitem{milo2002network}
Milo R, Shen-Orr S, Itzkovitz S, Kashtan N, Chklovskii D and Alon U 2002 {\em
  Science Signalling\/} {\bf 298} 824

\bibitem{cinquin2002roles}
Cinquin O and Demongeot J 2002 {\em Comptes rendus biologies\/} {\bf 325}
  1085--1095

\bibitem{liu2011controllability}
Liu Y~Y, Slotine J~J and Barab{\'a}si A~L 2011 {\em Nature\/} {\bf 473}
  167--173

\bibitem{wu2009identification}
Wu Y, Zhang X, Yu J and Ouyang Q 2009 {\em PLoS computational biology\/} {\bf
  5} e1000442

\bibitem{li2014quantifying}
Li G~W, Burkhardt D, Gross C and Weissman J~S 2014 {\em Cell\/} {\bf 157}
  624--635

\bibitem{russell1995energetics}
Russell J~B and Cook G~M 1995 {\em Microbiological reviews\/} {\bf 59} 48--62

\bibitem{buttgereit1995hierarchy}
Buttgereit F and Brand M~D 1995 {\em Biochem. J\/} {\bf 312} 163--167

\bibitem{aral2015coherence}
Aral N and Kabak{\c{c}}{\i}o{\u{g}}lu A 2015 {\em Physical Biology\/} {\bf 12}
  036002

\bibitem{ramo2005stability}
R{\"a}m{\"o} P, Kesseli J and Yli-Harja O 2005 {\em Chaos: An Interdisciplinary
  Journal of Nonlinear Science\/} {\bf 15} 034101--034101

\bibitem{wagner2005energy}
Wagner A 2005 {\em Molecular biology and evolution\/} {\bf 22} 1365--1374

\bibitem{willadsen2008understanding}
Willadsen K, Triesch J and Wiles J 2008 Understanding robustness in random
  boolean networks. {\em ALIFE\/} pp 694--701

\bibitem{alon2006introduction}
Alon U 2006 {\em An introduction to systems biology: design principles of
  biological circuits\/} (CRC press)

\bibitem{alon2007network}
Alon U 2007 {\em Nature Reviews Genetics\/} {\bf 8} 450--461

\bibitem{huang2005cell}
Huang S, Eichler G, Bar-Yam Y and Ingber D~E 2005 {\em Physical review
  letters\/} {\bf 94} 128701

\bibitem{luscombe2004genomic}
Luscombe N~M, Babu M~M, Yu H, Snyder M, Teichmann S~A and Gerstein M 2004 {\em
  Nature\/} {\bf 431} 308--312

\bibitem{davidich2008boolean}
Davidich M~I and Bornholdt S 2008 {\em PLoS One\/} {\bf 3} e1672

\bibitem{faure2006dynamical}
Faur{\'e} A, Naldi A, Chaouiya C and Thieffry D 2006 {\em Bioinformatics\/}
  {\bf 22} e124--e131

\bibitem{boole1}
Faur{\'e} A and Thieffry D 2009 {\em Mol. BioSyst.\/} {\bf 5} 1569--1581

\bibitem{boole2}
Lau K~Y, Ganguli S and Tang C 2007 {\em Physical Review E\/} {\bf 75} 051907

\bibitem{petri1}
Murata T 1989 {\em Proceedings of the IEEE\/} {\bf 77} 541--580

\bibitem{petri2}
Chaouiya C, Remy E, Ruet P and Thieffry D 2004 {\em Applications and Theory of
  Petri Nets 2004\/}  137--156

\bibitem{petri3}
Karlebach G and Shamir R 2008 {\em Nature Reviews Molecular Cell Biology\/}
  {\bf 9} 770--780

\bibitem{dif1}
Chen K~C, Wang T~Y, Tseng H~H, Huang C~Y~F and Kao C~Y 2005 {\em
  Bioinformatics\/} {\bf 21} 2883--2890

\bibitem{dif2}
Chen T, He H~L, Church G~M {\em et~al.\/} 1999 Modeling gene expression with
  differential equations {\em Pacific symposium on biocomputing\/} vol~4 p~4

\bibitem{li2004yeast}
Li F, Long T, Lu Y, Ouyang Q and Tang C 2004 {\em Proceedings of the National
  Academy of Sciences of the United States of America\/} {\bf 101} 4781--4786

\bibitem{chen2004integrative}
Chen K~C, Calzone L, Csikasz-Nagy A, Cross F~R, Novak B and Tyson J~J 2004 {\em
  Molecular biology of the cell\/} {\bf 15} 3841--3862

\bibitem{davidich2008differential}
Davidich M and Bornholdt S 2008 {\em arXiv preprint arXiv:0807.1013\/}

\bibitem{mendoza1999genetic}
Mendoza L, Thieffry D and Alvarez-Buylla E~R 1999 {\em Bioinformatics\/} {\bf
  15} 593--606

\bibitem{albert2003topology}
Albert R and Othmer H~G 2003 {\em Journal of theoretical biology\/} {\bf 223}
  1--18

\bibitem{gursky2001gap}
Gursky V~V, Reinitz J and Samsonov A~M 2001 {\em Chaos: An Interdisciplinary
  Journal of Nonlinear Science\/} {\bf 11} 132--141

\bibitem{leloup1998model}
Leloup J~C and Goldbeter A 1998 {\em Journal of biological rhythms\/} {\bf 13}
  70--87

\bibitem{akman2012digital}
Akman O~E, Watterson S, Parton A, Binns N, Millar A~J and Ghazal P 2012 {\em
  Journal of The Royal Society Interface\/} {\bf 9} 2365--2382

\bibitem{mendoza1998dynamics}
Mendoza L and Alvarez-Buylla E~R 1998 {\em Journal of theoretical biology\/}
  {\bf 193} 307--319

\bibitem{krumsiek2011hierarchical}
Krumsiek J, Marr C, Schroeder T and Theis F~J 2011 {\em PloS one\/} {\bf 6}
  e22649

\bibitem{remy2006logical}
Remy E, Ruet P, Mendoza L, Thieffry D and Chaouiya C 2006 {\em Transactions on
  Computational Systems Biology VII\/}  56--72

\bibitem{remy2008minimal}
Remy E and Ruet P 2008 {\em Bioinformatics\/} {\bf 24} i220--i226

\bibitem{fisher1925statistical}
Fisher R~A 1925 {\em Statistical methods for research workers\/} (Genesis
  Publishing Pvt Ltd)

\bibitem{morris1999hebb}
Morris R 1999 {\em Brain research bulletin\/} {\bf 50} 437

\bibitem{shou2011measuring}
Shou C, Bhardwaj N, Lam H~Y, Yan K~K, Kim P~M, Snyder M and Gerstein M~B 2011
  {\em PLoS Comput Biol\/} {\bf 7} e1001050--e1001050

\bibitem{lau2007function}
Lau K~Y, Ganguli S and Tang C 2007 {\em Physical Review E\/} {\bf 75} 051907

\end{thebibliography}

\end{document}